\documentclass[runningheads]{llncs}
\usepackage[T1]{fontenc}
\usepackage{graphicx}

\usepackage{url}
\usepackage{caption}
\usepackage{subcaption}
\usepackage{breqn}
\usepackage{amsfonts} 
\usepackage{relsize}
\begin{document}
%

 
\title{How~we~won~BraTS~2023~Adult~Glioma~challenge? Just faking it! \\ \large Enhanced Synthetic Data Augmentation and Model Ensemble for brain tumour segmentation \thanks{Supported by organizations University of Minho and Institute for Artificial Intelligence in Medicine}}
\titlerunning{Faking it!}
%
\author{André Ferreira \inst{1,2,3}\orcidID{0000-0002-9332-0091} \and
Naida Solak \inst{2,6} \and
Jianning Li \inst{2,3,4} \and
Philipp Dammann \inst{7} \and
Jens Kleesiek\inst{3,4,5} \and
Victor Alves\orcidID{0000-0003-1819-7051}\inst{1} \and
Jan Egger\inst{2,3,4,6}
}
\authorrunning{A. Ferreira et al.}
%
\institute{Center ALGORITMI / LASI, University of Minho, Braga, 4710-057,  Portugal \and
Computer Algorithms for Medicine Laboratory, Graz, Austria
\and
Institute for AI in Medicine (IKIM), University Medicine Essen, Girardetstraße 2, Essen, 45131, Germany
\and Cancer Research Center Cologne Essen (CCCE), University Medicine Essen, Hufelandstraße 55, Essen, 45147, Germany
\and German Cancer Consortium (DKTK), Partner Site Essen, Hufelandstraße 55, Essen, 45147, Germany
\and Institute of Computer Graphics and Vision, Graz University of Technology, Inffeldgasse 16, Graz, 8010, Austria
\and Department of Neurosurgery and Spine Surgery, University Hospital Essen, Essen, Germany\\
\email{\{id10656\}@alunos.uminho.pt}}
\maketitle              

\begin{abstract}
Deep Learning is the state-of-the-art technology for segmenting brain tumours. However, this requires a lot of high-quality data, which is difficult to obtain, especially in the medical field. Therefore, our solutions address this problem by using unconventional mechanisms for data augmentation. Generative adversarial networks and registration are used to massively increase the amount of available samples for training three different deep learning models for brain tumour segmentation, the first task of the BraTS2023 challenge. The first model is the standard nnU-Net, the second is the Swin UNETR and the third is the winning solution of the BraTS 2021 Challenge. The entire pipeline is built on the nnU-Net implementation, except for the generation of the synthetic data. The use of convolutional algorithms and transformers is able to fill each other's knowledge gaps. Using the new metric, our best solution achieves the dice results 0.9005, 0.8673, 0.8509 and HD95 14.940, 14.467, 17.699 (whole tumour, tumour core and enhancing tumour) in the validation set.

\end{abstract}
\section{Introduction}
Brain tumours originate from different cell types, mainly from glialcells (astrocytes, oligodendrocytes, microglia, ependymal cells) and are then referred to as gliomas. The World Health Organization (WHO) classifies brain tumours into grades 1 to 4 based on histologic features and molecular parameters. Grade 1 tumours are typically slow-growing and benign, and grade 4 tumours, such as glioblastomas (GBMs), are the most aggressive and malignant forms. Indeed, Glioblastomas are among the most deadly types of cancer due to their location and invasive growth. Patients diagnosed with glioblastoma now have a median survival of approximately 16 months with standard treatment (radiotherapy and temozolomide). Despite extensive research to improve diagnosis, characterization and treatment, the mortality rate of GBMs remains high and significant improvements in patient survival have been elusive.
Extensive research to improve diagnosis, characterisation and treatment has reduced the mortality rate of this disease \cite{menze2014multimodal}. Glioma segmentation is a critical step for tumour evolution, treatment efficacy assessment, survival prediction and treatment planning.
Multiple modalities of MRI scans (T1, T2, T1Gd and FLAIR) are usually used for accurate segmentation of the tumour and individual regions \cite{visser2019inter}.

MRI is a medical imaging technique that is often used to detect and assess response to glioma treatment \cite{vollmuth2019automated}. The development of new therapies for treatment depends on accurate segmentation. Manual segmentation has been used for this purpose, but is very time-consuming and suffers from inter- and intra-examiner variability \cite{egger2013gbm,visser2019inter}. Efforts have therefore been made to automate this process. Machine learning techniques have been the most advanced methods for performing such segmentation, but they have the disadvantage of requiring large, high-quality datasets for the training process in order to achieve the performance required for clinical purposes \cite{egger2022medical}.
 
The Brain Tumor Segmentation Challenge (BraTS) \cite{baid2021rsna,menze2014multimodal,bakas2017advancing,bakas2017segmentation_GBM,bakas2017segmentation_LGG}  provides a large, fully annotated and publicly available dataset for model development and promotes a competition to evaluate the latest state-of-the-art approaches for brain diffuse glioma segmentation. This competition was launched in 2012 and continues to evolve each year, adding more samples and many different tasks. The 2023 competition includes 9 different tasks, the first of which is the traditional segmentation of adult gliomas.

\subsection{State-of-the-art}
Since the challenge of 2014, deep convolutional networks have been the state-of-the-art for brain tumour segmentation \cite{urban2014multi,pereira2016brain}.
Recently, the most advanced strategies are mainly based on deep neural networks (DNNs), due to the rapid development of these tools, the availability of increasingly powerful GPUs and the availability of training data. Most solutions are based on the U-Net \cite{ronneberger2015u} which has yielded convincing results. Many architectural changes to the U-Net have been introduced to improve it, e.g., residual connections\cite{jiang2020two},  densely connected layers \cite{mckinley2019ensembles,zhao2020bag} and attention mechanisms \cite{hatamizadeh2021swin}.

The winners of the last 6 editions all used DNNs. Kamnitsas et al. \cite{kamnitsas2018ensembles} (2017 winner) explore the ensemble of multiple models and architecture (EMMA), more specifically the 3D convolutional networks DeepMedic \cite{kamnitsas2017efficient,kamnitsas2015multi}, FCN \cite{long2015fully} and U-Net \cite{ronneberger2015u}. The use of EMMA seems to reduce the influence of the meta-parameters of each model and helps to avoid overfitting. 

The winners of the 2020 edition \cite{isensee2021nnu_brats}, with respect to task 1, propose the use of the nnU-Net \cite{isensee2021nnu}, a U-Net based architecture, as a baseline and implements some BraTS specific optimisations. These optimizations are: optimising regions rather than individual classes, using a bigger batch size (from 2 to 5), applying a more aggressive data augmentation, replacing the instance normalisation with batch normalisation (which is better with more aggressive data augmentation), using of batch dice instead of regular dice, and applying post-processing distinct from the regular nnU-Net.

The winners of 2021 edition \cite{luu2021extending} also use the nnU-Net as baseline, using the same BraTS-specific optimizations as \cite{isensee2021nnu} and new optimizations. They claim that due to the change in the dataset from the 2020 edition (494 cases) to the 2021 edition (1470 cases), they decided to use a larger network by doubling the number of filters in the encoder part of the nnU-Net, and increasing the maximum number of filter in the bottleneck to 512, while keeping the decoder intact. Batch normalisation is replaced by group normalisation as it performs better with small batch sizes, allowing the use of batch size 2 instead of 5. Axial attention decoder was also applied but not tested in the final phase of the BraTS challenge, and it did not improve the results in the 5-fold cross validation. 

The winners of the 2022 edition \cite{zeineldin2022multimodal} archive the best results by using an ensemble of three different architectures:  DeepSeg \cite{zeineldin2020deepseg}, the improved nnU-Net proposed by \cite{luu2021extending}, and DeepSCAN \cite{mckinley2019ensembles}. The ensemble is created using the Simultaneous Truth and Performance Level Estimation (STAPLE). 2023 Challenge \textit{BraTS-Africa} \cite{adewole2023brain} also uses the STAPLE ensemble of three different models to create the ground truth segmentations.

It is important to note that the same post-processing is done for all three solutions. As explained in detail in \cite{isensee2021nnu_brats}, there is ground truth without an ET label, so the BraTS evaluation gives the worst possible results for false positive predictions. To mitigate this scenario, when the number of voxels is below a certain threshold, ET is replaced by NCR. A threshold of 200 is used in all three experiments. This year (BraTS2023), however, a new metric is used that requires new post-processing techniques, as will be explained later.

Our approach consists of solving the problem of brain tumour segmentation by increasing the amount of available data, using two different, non-conventional strategies for data augmentation. Furthermore, convolutional neural networks (CNNs) and transformer-based networks are assumed to complement each other, which is why ensemble of these two distinct architectures are also tested.

\section{Methods}
The machine used for all these tasks is a IKIM cluster node with 6 NVIDIA RTX 6000, 48 GB of VRAM, 1024 GB of RAM, and AMD EPYC 7402 24-Core Processor.

\subsection{Data}
\begin{figure}[h!]
     \centering
     \begin{subfigure}[b]{0.19\textwidth}
         \centering
         \includegraphics[width=\textwidth]{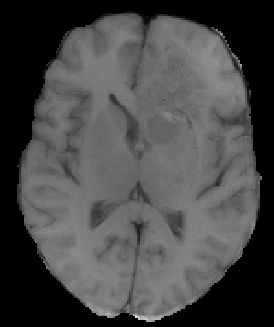}
         \caption{T1}
         \label{fig:T1}
     \end{subfigure}
     \hfill
     \begin{subfigure}[b]{0.19\textwidth}
         \centering
         \includegraphics[width=\textwidth]{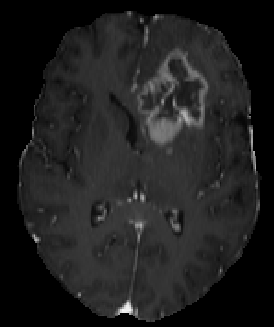}
         \caption{T1Gd}
         \label{fig:T1Gd}
     \end{subfigure}
     \hfill
     \begin{subfigure}[b]{0.19\textwidth}
         \centering
         \includegraphics[width=\textwidth]{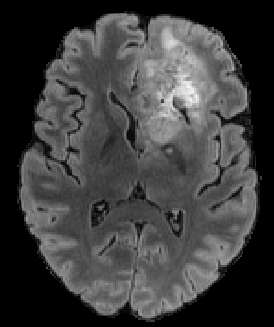}
         \caption{FLAIR}
         \label{fig:FLAIR}
     \end{subfigure}
     \hfill
     \begin{subfigure}[b]{0.19\textwidth}
         \centering
         \includegraphics[width=\textwidth]{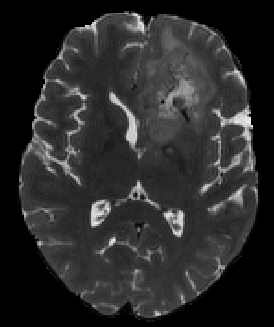}
         \caption{T2}
         \label{fig:T2}
     \end{subfigure}
     \hfill
     \begin{subfigure}[b]{0.19\textwidth}
         \centering
         \includegraphics[width=\textwidth]{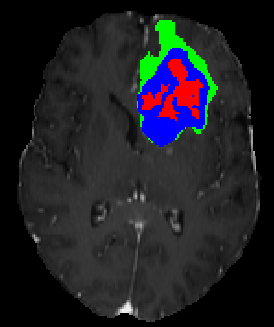}
         \caption{T1Gd (label)}
         \label{fig:t1ce_seg}
     \end{subfigure}
     \hfill
        \caption{All four modalities of the sample BraTS-GLI-00000-000 and the corresponding groun-truth from the BraTS2023 dataset.}
        \label{fig:BraTS-GLI-00000-000}
\end{figure}

The Task 1 dataset consists of 1470 patients, each of which contains 4 modalities, as can be seen in Figure \ref{fig:BraTS-GLI-00000-000} (T1, T1Gd, T2 and FLAIR). 1251 cases have the corresponding ground truth, so this subset is used for training, and the remaining (219 cases) that do not contain a freely available ground truth form the subset for evaluation. The test set in this year's (2023) edition contains many more routine clinical mpMRI scans. All MRI scans were pre-processed as follows: were co-registered to the same anatomical template, interpolated to the same resolution (1 mm3), and skull-striped. The scans have the shape 240×240×155.

The ground truth of the subset of evaluation is hidden from the participants, being only possible to access the Dice scores and 95\% Hausdorff distance through the participation platform. The evaluation is performed by sub-region and not by individual label. The sub-regions are the enhancing tumour (ET), the tumour core (TC), and the whole tumour (WT). This year (BraTS2023), the value 4 was replaced by the value 3. The naming convention has also changed from the previous challenges.

\subsection{Data augmentation}

\subsubsection{Data augmentation via registration:}
\label{subsub:Registration}
Inspired by the winning solution for the AutoImplant Challenge 2020 \cite{ellis2020deep}, a larger dataset was build via registration. Each scan can be registered with any other scan and then warped into the other. Advanced Normalization Tools (ANTs \footnote{http://stnava.github.io/ANTs/}) package is used to perform this registration. It is a software package used for normalizing data to a template, and it provides different scripts
(such as \textit{antsRegistrationSyNQuick.sh}) that enable applying different transformations on the images, such as rigid, affine, non-linear and all of them combined. Usually, one (first) image is used as the moving image, and the other (second) one as the fixed image - meaning that the moving image is warped into the fixed image space by applying the computed transformation, and also the inverse transformation in order to warp the fixed image into the moving image space.

\begin{figure}[h!]
     \centering
    
     \begin{subfigure}[b]{0.19\textwidth}
         \centering
         \includegraphics[width=\textwidth]{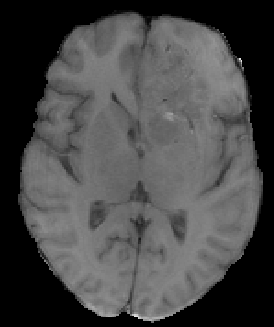}
         \caption{T1}
         \label{fig:Naida_01_T1}
     \end{subfigure}
     \hfill
     \begin{subfigure}[b]{0.19\textwidth}
         \centering
         \includegraphics[width=\textwidth]{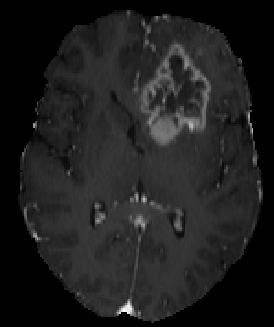}
         \caption{T1Gd}
         \label{fig:Naida_01_T1Gd}
     \end{subfigure}
     \hfill
     \begin{subfigure}[b]{0.19\textwidth}
         \centering
         \includegraphics[width=\textwidth]{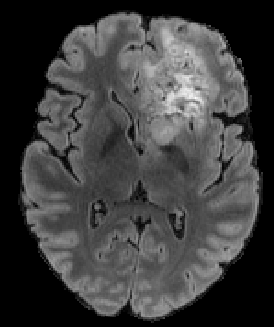}
         \caption{FLAIR}
         \label{fig:Naida_01_FLAIR}
     \end{subfigure}
     \hfill
     \begin{subfigure}[b]{0.19\textwidth}
         \centering
         \includegraphics[width=\textwidth]{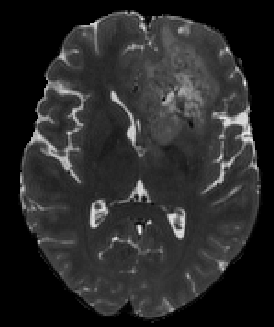}
         \caption{T2}
         \label{fig:Naida_01_T2}
     \end{subfigure}
     \hfill
     \begin{subfigure}[b]{0.19\textwidth}
         \centering
         \includegraphics[width=\textwidth]{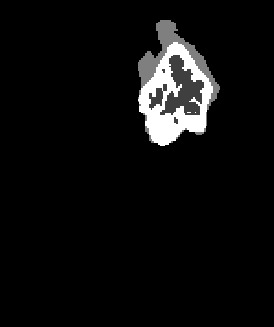}
         \caption{Label}
         \label{fig:Naida_01_seg}
     \end{subfigure}
     \begin{subfigure}[b]{0.19\textwidth}
         \centering
         \includegraphics[width=\textwidth]{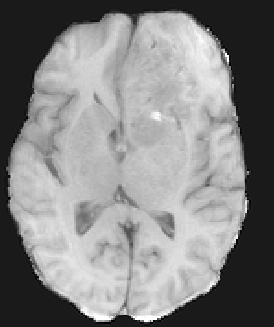}
         \caption{T1}
         \label{fig:Naida_02_T1}
     \end{subfigure}
     \hfill
     \begin{subfigure}[b]{0.19\textwidth}
         \centering
         \includegraphics[width=\textwidth]{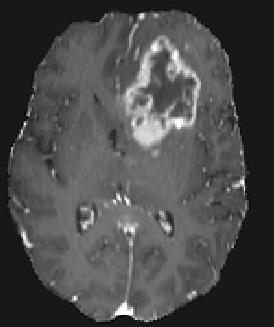}
         \caption{T1Gd}
         \label{fig:Naida_02_T1Gd}
     \end{subfigure}
     \hfill
     \begin{subfigure}[b]{0.19\textwidth}
         \centering
         \includegraphics[width=\textwidth]{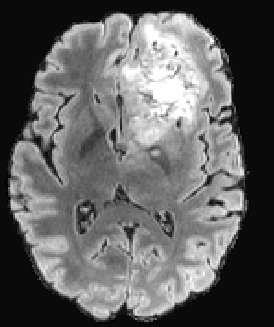}
         \caption{FLAIR}
         \label{fig:Naida_02_FLAIR}
     \end{subfigure}
     \hfill
     \begin{subfigure}[b]{0.19\textwidth}
         \centering
         \includegraphics[width=\textwidth]{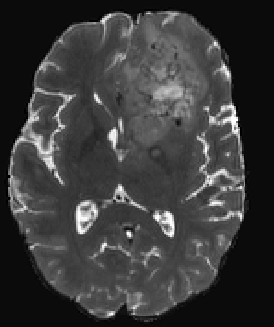}
         \caption{T2}
         \label{fig:Naida_02_T2}
     \end{subfigure}
     \hfill
     \begin{subfigure}[b]{0.19\textwidth}
         \centering
         \includegraphics[width=\textwidth]{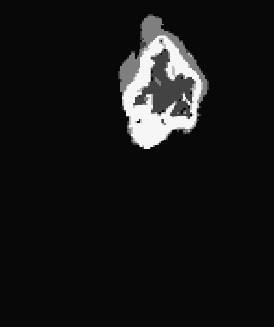}
         \caption{Label}
         \label{fig:Naida_02_seg}
     \end{subfigure}
    
     \begin{subfigure}[b]{0.19\textwidth}
         \centering
         \includegraphics[width=\textwidth]{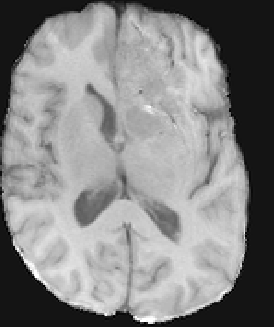}
         \caption{T1}
         \label{fig:Naida_03_T1}
     \end{subfigure}
     \hfill
     \begin{subfigure}[b]{0.19\textwidth}
         \centering
         \includegraphics[width=\textwidth]{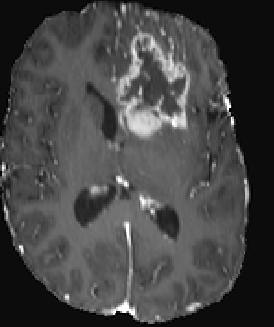}
         \caption{T1Gd}
         \label{fig:Naida_03_T1Gd}
     \end{subfigure}
     \hfill
     \begin{subfigure}[b]{0.19\textwidth}
         \centering
         \includegraphics[width=\textwidth]{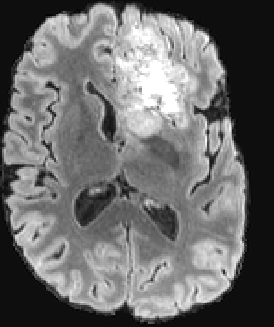}
         \caption{FLAIR}
         \label{fig:Naida_03_FLAIR}
     \end{subfigure}
     \hfill
     \begin{subfigure}[b]{0.19\textwidth}
         \centering
         \includegraphics[width=\textwidth]{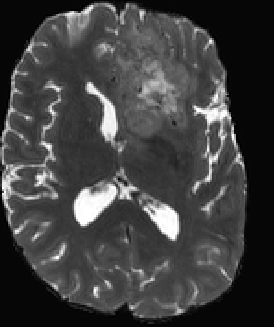}
         \caption{T2}
         \label{fig:Naida_03_T2}
     \end{subfigure}
     \hfill
     \begin{subfigure}[b]{0.19\textwidth}
         \centering
         \includegraphics[width=\textwidth]{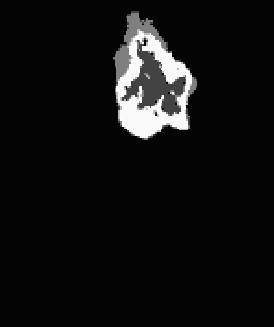}
         \caption{Label}
         \label{fig:Naida_03_seg}
     \end{subfigure}
        \caption{All four modalities of the sample BraTS-GLI-00000-000 (first row) and respective ground-truth registered with the samples BraTS-GLI-00002-000 (second row) and BraTS-GLI-00003-000 (last row) from the BraTS2023 dataset.}
        \label{fig:Registration_samples}
\end{figure}

Only the training set was used to create new samples via registration, since it is the only one that contains the ground truth. The transformation and inverse transformation matrix are computed for each case and applied to each scan (including the ground truth). After creating a reasonable amount of registered data (23049 samples, 92000 MRI scans and 23049 ground truths), all data were converted to integer, as the results of registering creates floats. This process took around 2 weeks. Figure \ref{fig:Registration_samples} presents a sample.

\subsubsection{Data augmentation with GANs:}
\label{subsub:GANs}
Generative adversarial networks (GANs) are known for their ability to generate realistic data. Ferreira et al. (2022) \cite{ferreira2022gan} presents an overview of the generation of realistic volumetric data. In this systematic review, it can be seen that a large number of works use GANs for various tasks, e.g. denoising, classification, segmentation, image translation, reconstruction and others. In many cases, synthetic data generated by GANs is even used to increase the amount of data available for training deep learning models, i.e. for data augmentation.

\begin{figure}[h!]
     \centering
     \includegraphics[width=\textwidth]{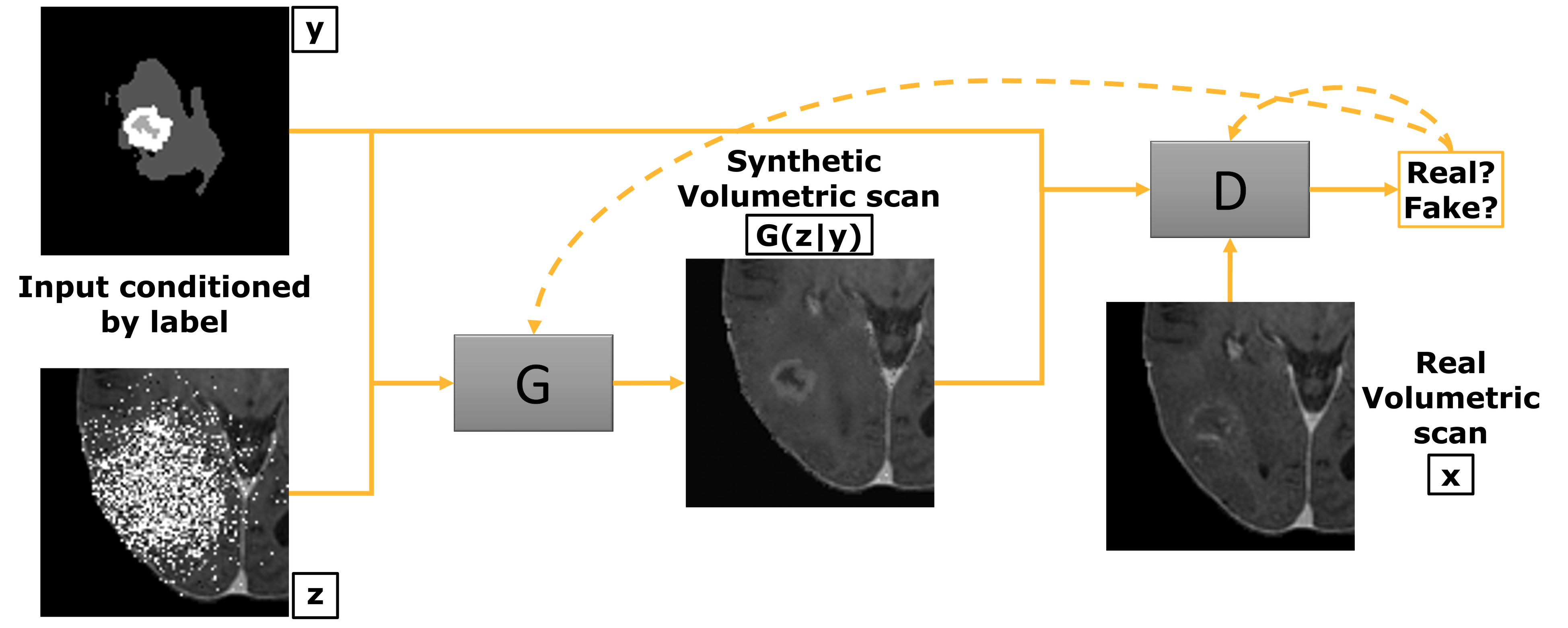}
        \caption{The training pipeline of the \textbf{GliGAN}. The noise scan (z) and the label (y) are concatenated and fed into the generator (G). The discriminator (D) assesses the realism of a real scan and the reconstruction.}
        \label{fig:GANs-train}
\end{figure}

In this work, a GAN (represented in Figure \ref{fig:GANs-train}, referred to as \textbf{GliGAN}) is trained to generate synthetic tumours that can be randomly inserted into the healthy parts of the brain. Placing synthetic tumours in the provided training set reduces the class imbalance between tumour labels and healthy brain tissue and creates greater variability in tumour properties and locations. The GAN architecture consists of a generator and a discriminator. The generator uses the MONAI\footnote{https://monai.io/} implementation of the  Swin UNETR \cite{hatamizadeh2021swin}, with parameters: img\_size=(96, 96, 96), in\_channels=4, out\_channels=1, and feature\_size=48. The discriminator is a CNN based on Ferreira et al. (2022) \cite{ferreira2022generation}, with increased number of layers (one more 3D convolution at the end, with stride 1, kernel size 3, padding 0 and no spectral normalisation), and sigmoid before the output. The learning rate of both the generator and the discriminator is 0.0001 with a batch size of 2. The generator is trained twice per iteration, while the discriminator is only trained once per iteration. 

The input of the generator is created as follows:
Each scan is cropped with centre in the tumour volume (size 96×96×96), normalised between [-1,1], added Gaussian noise (mean 0 and standard deviation 1) and normalised between [-1,1] again, resulting in a noisy scan as can be seen in Figure \ref{fig:GANs-train} (z). Voxels with a tumour label different from 0 are replaced by Gaussian noise. Neighbouring voxels may be selectively replaced by noise, contingent on a given probability. The probability of the voxel value being replaced decreases with distance from the tumour centre, producing a spherical effect. This probably is computed taking into account the size of the tumour. As bigger tumours might have a greater impact on the surrounding tissue, more distant voxels will have more probability of being replaced by noise than in cases with smaller tumours. It was decided that the probability would decay linearly with the distance to the centre, following the equation \ref{eq:prob}.

\begin{dmath}
\label{eq:prob}
 prob = \frac{83}{exponent^{distance}+82}
\end{dmath}
where the $exponent$ is defined by equation \ref{eq:exp}
\begin{dmath}
\label{eq:exp}
exponent=-\frac{0.2}{68}*max\_size + 1.1-96*-\frac{0.2}{68}
\end{dmath}
and $max\_size$ is the greater size among each three planes.

\begin{figure}[h!]
     \centering
     \begin{subfigure}[b]{0.25\textwidth}
         \centering
         \includegraphics[width=\textwidth]{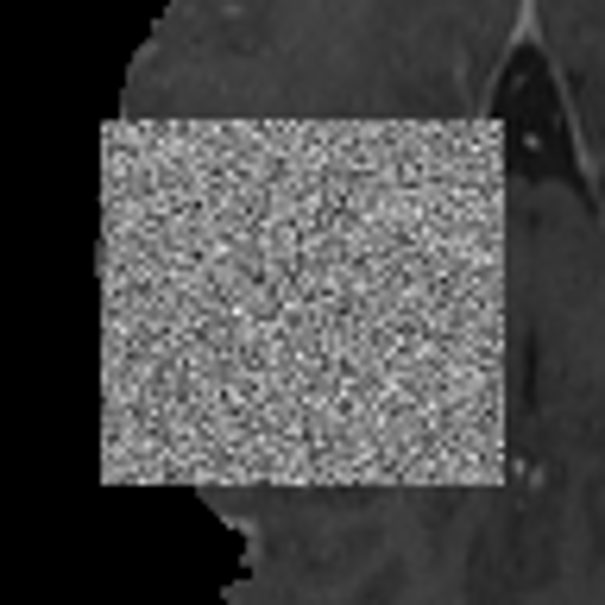}
         \caption{}
         \label{fig:square_noise}
     \end{subfigure}
     \begin{subfigure}[b]{0.25\textwidth}
         \centering
         \includegraphics[width=\textwidth]{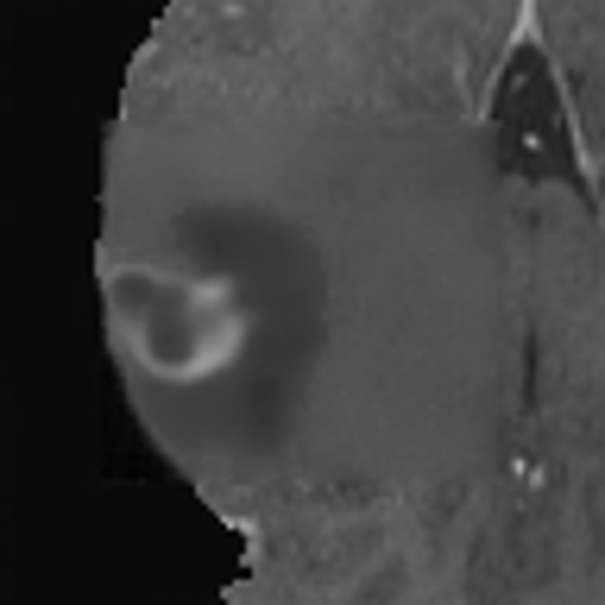}
         \caption{}
         \label{fig:square_recons}
     \end{subfigure}
    \caption{Using a square shaped noise: a) Crop of a real scan (T1Gd) with squared noise b) Reconstruction using GANs}
    \label{fig:square_noise_all}
\end{figure}

The tumour centre in calculated considering the first and last slice where the tumour appears in all three planes.  This strategy allows for more realistic generation than only adding noise to the tumour or placing a cube of noise, as it is harder to detect the edges of the noise. Only replacing the tumour voxels by noise would not allow the network to learn how to replicate the mass effect of the growing tumour. Square noise has been tested but produces unrealistic results, as shown in Figure \ref{fig:square_noise_all}.

The loss functions of the generator (G) and discriminator (D) are defined by equations \ref{eq:G_1} and \ref{eq:D}:

\begin{dmath}
\label{eq:G_1}
L_{G}=-\lambda_{1}\mathbb{E}{_{y,z}}[\log D(G(z|y))]+\lambda_{2}\mathbb{E}{_{x,y,z}}\left \| x-G(z|y)) \right \|_{MAE}
\end{dmath}

\begin{dmath}
\label{eq:D}
L_{D}=\mathbb{E}{_{y,z}}[\log D(G(z|y))]-\mathbb{E}{_{x,y}}[\log D(x|y)]
\end{dmath}

The training is divided into two steps. The first step uses $\lambda_{1}=1$ and $\lambda_{2}=5$ for 200000 iterations. After performing the first step, it was found that the tissue visible to the generator (voxels without noise) was noisy, but the tumour volume was realistic. To solve this problem, the network was trained for another 1000 epochs, linearly increasing the weight of the mean absolute error (MAE) component ($\lambda_{2}$) and reducing the weight of the adversarial loss ($\lambda_{1}$), where $\lambda_{1} = \frac{1}{\lambda_{2}}$, ending with $\lambda_{2}=100$, i.e., $\lambda_{2}=\frac{100-1}{1000}*epoch+1$.

\begin{figure}[h!]
     \centering
     \begin{subfigure}[b]{0.24\textwidth}
         \centering
         \includegraphics[width=\textwidth]{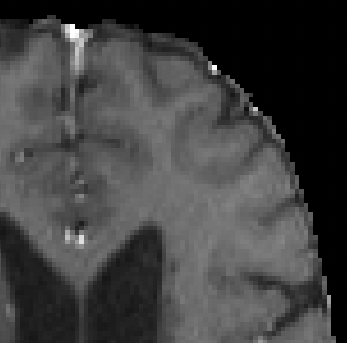}
         \caption{}
         \label{fig:healthy}
     \end{subfigure}
     \hfill
     \begin{subfigure}[b]{0.24\textwidth}
         \centering
         \includegraphics[width=\textwidth]{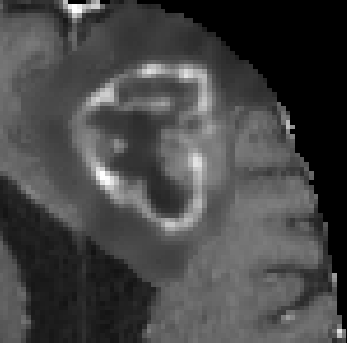}
         \caption{}
         \label{fig:baseline}
     \end{subfigure}
     \hfill
     \begin{subfigure}[b]{0.24\textwidth}
         \centering
         \includegraphics[width=\textwidth]{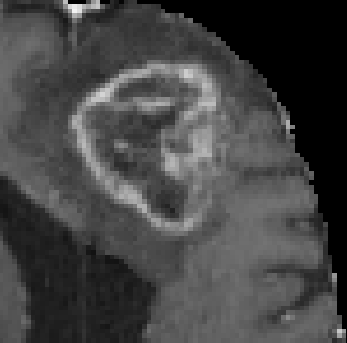}
         \caption{}
         \label{fig:1_step_GAN}
     \end{subfigure}
     \hfill
     \begin{subfigure}[b]{0.24\textwidth}
         \centering
         \includegraphics[width=\textwidth]{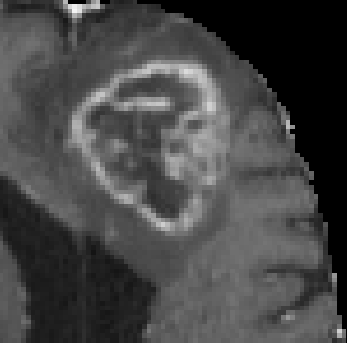}
         \caption{}
         \label{fig:2_step_GAN}
     \end{subfigure}
    \caption{Differences between a) Healthy crop b) baseline c) first step training d) second step training.}
    \label{fig:synthetic_noise_GANs}
\end{figure}

This allowed for a realistic surrounding tissue and tumour with realistic texture and overall appearance, as shown in Figure \ref{fig:synthetic_noise_GANs} (second row). The baseline uses only the MAE component as the loss function.

For the generation process (the new scans with the fake tumour for the segmentation task), two approaches were tested to create two datasets. In the first datasets, for each real case in the training dataset, 30 random labels (from the remaining dataset) were selected and randomly placed in a healthy part of the scan. From the total set, 23049 cases were randomly selected. This dataset is referred to as \textbf{G}. The second dataset was created using a random label generator, i.e. another GAN based on \cite{ferreira2022generation} was trained to generate new synthetic labels. For training this GAN, all labels smaller or equal to 96×96×96 were cropped and resized to 64×64×64. The synthetic labels are then used as input to the \textbf{GliGAN}. This dataset is referred to as \textbf{rG}. To allow a fair comparison between all data augmentation strategies, 23049 cases were generated for each strategy.

\subsubsection{Pre-procesing:}
Pre-processing is performed by the nnU-Net pipeline. Before the training step, the brain voxels of each scan are normalised using z-score normalisation, keeping the background at zero.

\subsubsection{Networks:}
\label{subsub:Networks}
Multiple networks were tested to determine which network (or ensemble) provided better results. Each network was implemented in the newer version of the nnU-Net\footnote{https://github.com/MIC-DKFZ/nnUNet} to take advantage of the pre-processing and data augmentation pipeline provided by this framework.

\begin{figure}[h!]
     \centering
     \includegraphics[width=\textwidth]{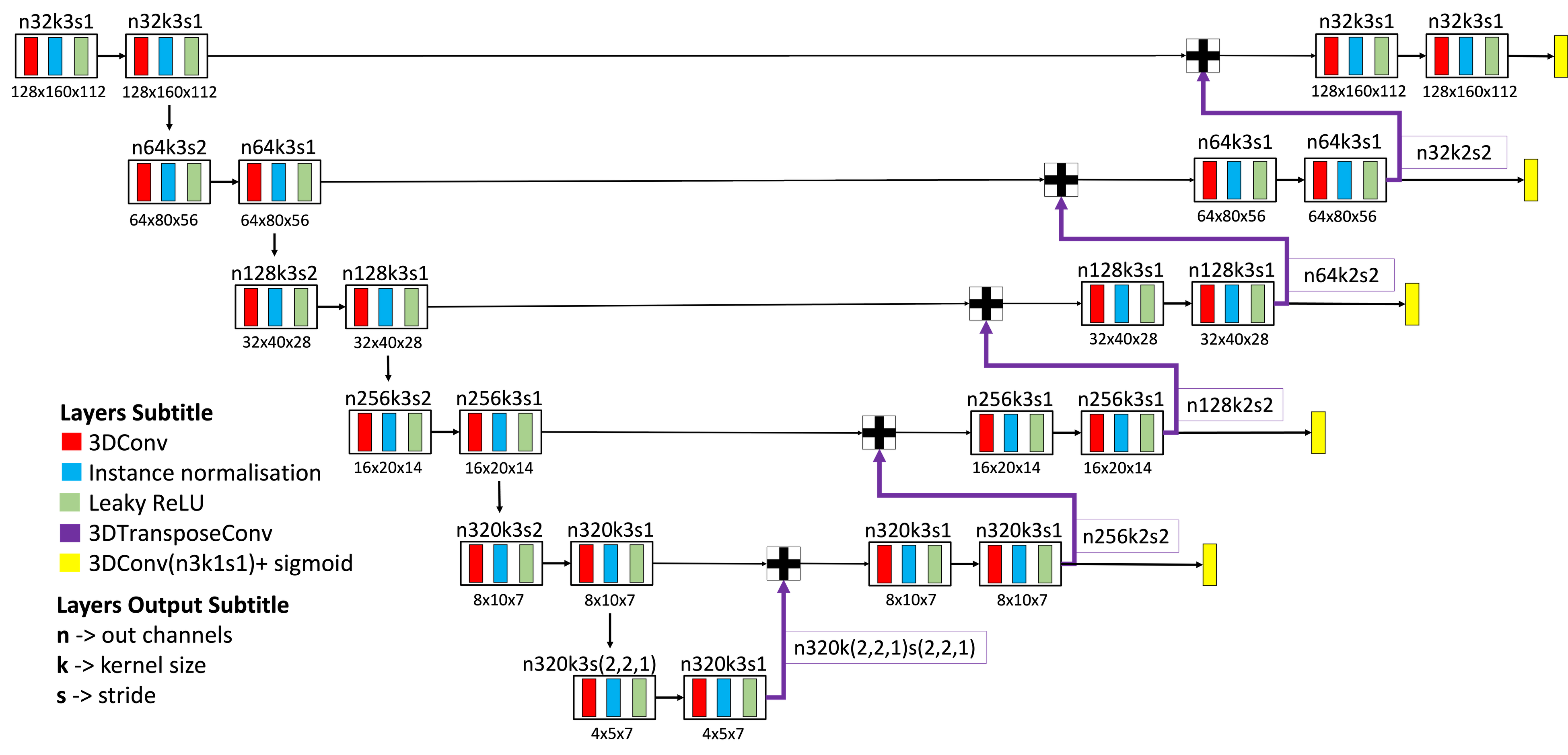}
    \caption{Baseline architecture.}
    \label{fig:baseline_arch}
\end{figure}

\paragraph{Baseline (B):} The fully automated framework nnU-Net \cite{isensee2021nnu} was used as a baseline (3D full resolution), without any configuration changes. Figure \ref{fig:baseline_arch} shows this architecture in detail. 
The input is random patches of the shape 128×160×112. Batch size 5, region-based training, and deep supervised.

\paragraph{Swin UNETR (S):}

The Swin UNETR \cite{hatamizadeh2021swin} is a U-Net like network in which the convolutional encoder is replaced by Swin transformer blocks. This encoder is, therefore, capable of capturing long-range information, as opposed to fully convolutional networks. The Swin Transformer uses shifted windows, allowing the use of high resolution images, as is the case of the BraTS dataset (by having linear computational complexity regarding the image size \cite{liu2021swin}).

The architecture of the Swin UNETR is described in \cite{hatamizadeh2021swin}. The input are random patches of shape 128×128×128, and the remaining pre-processing, including regular data augmentation, is applied by the nnU-Net pipeline. Since this network is heavier than the nnU-Net, a batch size of 4 is used. Deep supervision is also used.

We hypothesise that a transformer-based architecture is complementary to the fully convolutional nnU-Net architecture, as found by \cite{zhou2021nnformer}.

\paragraph{2021 winner (L):}

\begin{figure}[h!]
     \centering
     \includegraphics[width=\textwidth]{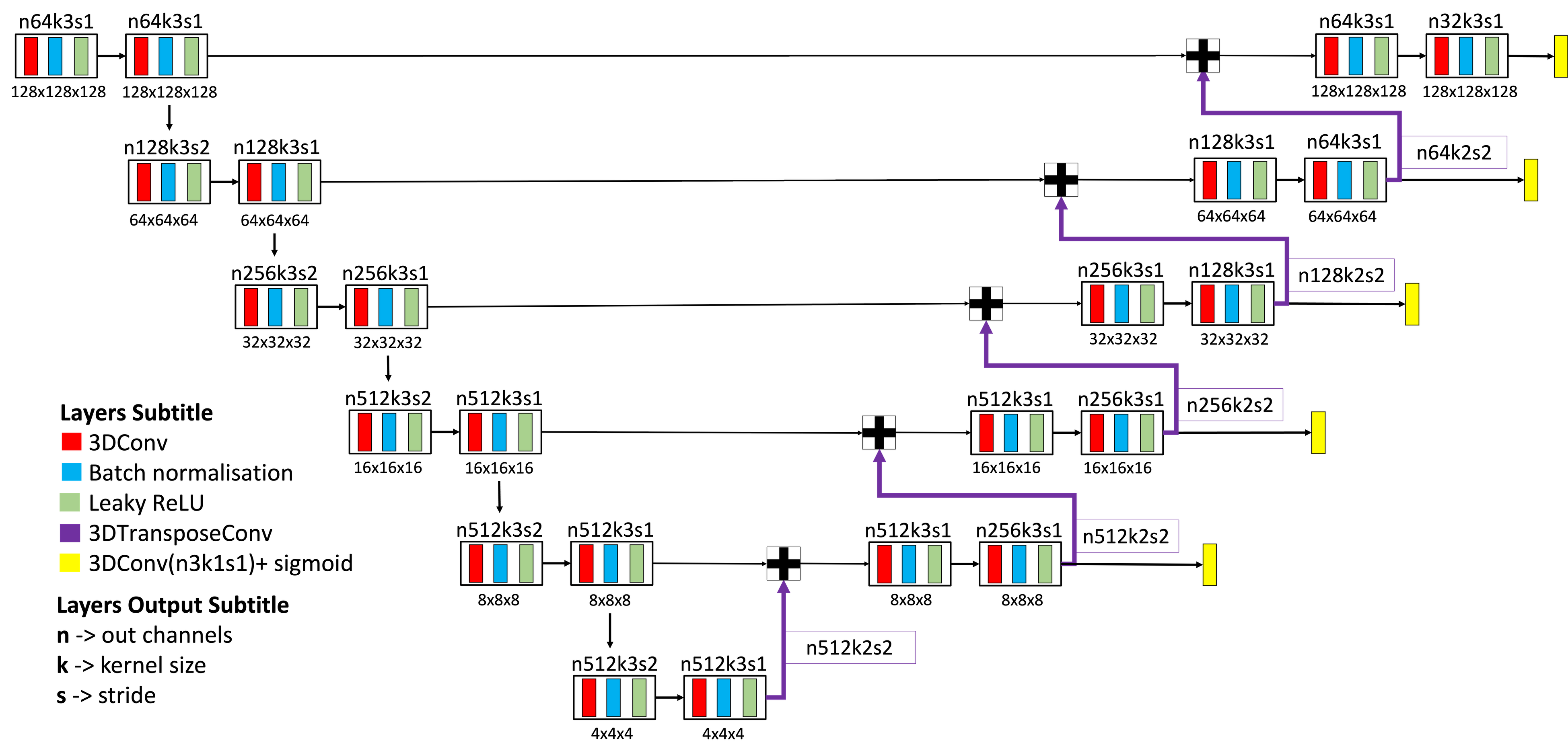}
    \caption{Architecture based on the winner of the 2021 edition of the BraTS challenge \cite{luu2021extending}.}
    \label{fig:winner2021_arch}
\end{figure}

The winners of the 2021 edition \cite{luu2021extending} use the framework provided by nnU-Net and implemented some improvements over the nnU-Net solution proposed in 2020 \cite{isensee2021nnu_brats}. Isensee et al. \cite{isensee2021nnu_brats} purposes the use of a more aggressive data augmentation than the provided by default in the nnU-Net pipeline, and therefore to use of batch normalisation instead of instance normalisation, as this seems to produce better results with a very aggressive data augmentation. Since the dataset used for training includes samples generated by our own data augmentation strategies as well as those explained in \cite{isensee2021nnu_brats}, we believe that batch normalisation gives better results than any other normalisation. Batch dice is also used for gradient computation instead of sample Dice \cite{isensee2021nnu_brats}.  In addition, Luu et al. \cite{luu2021extending} double the number of filters in the encoder and increases the size of the bottleneck to 512, as can be seen in Figure \ref{fig:winner2021_arch}.  
They also use group normalisation as they claim it is better for small batch sizes. However, we were able to use a batch size of 5, so batch normalisation would be the best option.

\subsection{Selection criterion}

As mentioned by Isensee et al. \cite{isensee2021nnu_brats}, the BraTS challenge follows a "rank and then aggregate" approach for ranking the proposed solutions, as it is well suited for the using of different segmentation metrics \footnote{https://zenodo.org/record/3718904}. This ranking method works as follows: each participant is ranked for each of the testing cases (X); each case includes 3 regions, the metrics used are the Dice similarity coefficient (DSC) and  the 95\% Hausdorff distance (HD95); this makes X×3×2; the final ranking is the average of all these rankings normalised by the number of participants, from 0 (best) to 1 (worst). In situations where the baseline data does not have a specific label, false positives are strongly penalised by assigning the worst possible value for each metric (DSC 0 and HD95 374) and the best possible result when the model outputs an empty label (DSC 1 and HD95 0).

This year (BraTS2023), two new performance metrics are used, the lesion-based Dice score and the lesion-based Hausdorff distance-95. With these metrics, the evaluation is done at the lesion level rather than at the scan level. These new metrics can be used to assess how well the model can detect multiple tumours in the same case. Therefore, for each case, the DSC and HD95 are calculated for each lesion and averaged per patient. Thus, the results are heavily penalised by the segmentation of non-existent tumours (FP) and the lack of segmentation (FN). FNs are almost impossible to avoid in a post-processing step, but FP can be reduced by using a suitable threshold to remove some segmentations, although this comes with a slight increase in the number of FNs.

Therefore, selecting the solution with the best DSC and/or HD95 is not the best approach for an optimal performance on the BraTS2023 competition. For this choice, an implementation based on the ranking strategy of the BraTS competition is used, in order to select the best solution.

\subsection{Ensemble strategy}
Sawant et al. \cite{sawant2023comparing} claims that the maximum success of ensembles is only achieved when the number of models used is between 60 and 70. However, training and inference of more than 60 models makes the segmentation task too time-consuming and computationally heavy. Therefore, it was decided to train one model for each architecture and for each data augmentation strategy, giving a total of 3×3=9 models. Each model was trained with a 5-fold cross-validation resampling method, i.e. a total of 9×5=45 checkpoints.

In order to get the best out of all the models and find the best solution, several ensemble methods were tested: 

\begin{itemize}
    \item Averaging: averaging the probabilities of all 45 checkpoints;
    \item STAPLE: Application of the STAPLE algorithm to each label provided by each network (after averaging the 5-fold probability map and converting to integers);
    \item CNN: Training a CNN to produce the final labels using the probability maps of each network as input (after averaging the 5-folds probability maps);
    \item Weighting: Similar to the previous method, but instead of convolutions, a learnable parameter is used that learns how much weight each region of each individual probability mask of each network should have.
\end{itemize}

\section{Results}

We refer to our solutions using the following abbreviations:
\begin{itemize}
    \item \textbf{B}: Baseline network, explained in section \ref{subsub:Networks}.
    \item \textbf{S}: Swin UNETR, explained in section \ref{subsub:Networks}.
    \item \textbf{L}: Architecture based on BraTS2021 winner, explained in section \ref{subsub:Networks}.
    \item \textbf{G}: Synthetic data generated by the GliGAN, explained in section \ref{subsub:GANs}.
    \item \textbf{rG}: Synthetic data generated by the random label generator and GliGAN, explained in section \ref{subsub:GANs}.
    \item \textbf{R}: Synthetic data generated by registration, explained in section \ref{subsub:Registration}.
\end{itemize}

Since, in this edition of BraTS a new metric was introduced that takes into account the number of tumours in the ground-truth and predictions, penalising the FPs and FNs, Table \ref{tab:5-fold} presents both legacy (old metric) and new metric for the training set. 
Each solution is formally defined as $S_M^{DA}$ where $S$ is the solution, $M$ the model and $DA$ the data augmentation strategy. E.g., the solution which uses the baseline network \textbf{B} and the synthetic dataset \textbf{G} is represented as $S_B^G$. For ensembles, several model and data represented are used, e.g.,$S_{B,S}^{G,rG}$.
Table \ref{tab:our_winner2022} shows the legacy results of our final submission in the validation set, the ensemble of each data augmentation strategy and the winners of 2021 \cite{luu2021extending} and 2022 \cite{zeineldin2022multimodal}.
Table \ref{tab:validation} shows the results (using the new metric) of the online validation platform. This evaluation is performed online as the participants have no access to the ground-truth. Only ensembles were submitted to the platform as they produced the better results and the number of submissions is limited. The results of the 3 model ensembles help to evaluate how good each data augmentation strategy is.

\begin{table}[]
\centering
\caption{Results of the training set. The best results are in \textbf{bold} and the second best \underline{underlined}. The "All" is defined as $S_{B,L,S,B,L,S,B,L,S}^{G,G,G,rG,rG,rG,R,R,R}$}
\label{tab:5-fold}
\begin{tabular}{|c|cccccccc|}
\hline
\textbf{} & \multicolumn{4}{c}{\textbf{Legacy (DSC)}} & \multicolumn{4}{c|}{\textbf{New metric (DSC)}} \\ \hline
Solutions & WT & TC & ET & \multicolumn{1}{c|}{Mean} & WT & TC & ET & Mean \\ \hline
$S_B^G$ & 0.9388 & 0.9204 & 0.8833 & \multicolumn{1}{c|}{0.9142} & 0.8326 & 0.8690 & 0.8168 & 0.8395 \\
$S_S^G$  & 0.9378 & 0.9148 & 0.8787 & \multicolumn{1}{c|}{0.9104} & 0.7855 & 0.8533 & 0.8010 & 0.8133 \\
$S_L^G$  & 0.9377 & 0.9183 & 0.8846 & \multicolumn{1}{c|}{0.9136} & 0.8332 & 0.8673  & 0.8207 & 0.8404 \\
$S_B^{rG}$  & 0.9405 & 0.9213 & 0.8819 & \multicolumn{1}{c|}{0.9146} & 0.8265 & 0.8774 & 0.8224 & 0.8421 \\
$S_S^{rG}$  & 0.9397 & 0.9165 & 0.8797 & \multicolumn{1}{c|}{0.9120} & 0.7627 & 0.8569 & 0.8057 & 0.8084 \\
$S_L^{rG}$  & 0.9400 & 0.9188 & \underline{0.8873} & \multicolumn{1}{c|}{0.9154} & 0.8525 & 0.8669 & 0.8160 & 0.8451 \\
$S_B^R$  & 0.9380 & 0.9180 & 0.8742 & \multicolumn{1}{c|}{0.9101} & 0.8423 & 0.8731 & 0.8118 & 0.8424 \\
$S_S^R$  & 0.9357 & 0.9085 & 0.8680 & \multicolumn{1}{c|}{0.9041} & 0.8013 & 0.8415 & 0.7889 & 0.8106 \\
$S_L^R$  & 0.9387 & 0.9139 & 0.8830 & \multicolumn{1}{c|}{0.9119} & 0.8401 & 0.8664 & 0.8183 & 0.8416 \\
$S_{B,L,S}^{G,G,G}$ & 0.9409 & 0.9211 & \textbf{0.8879} & \multicolumn{1}{c|}{0.9166} & 0.8526 & 0.8742 & 0.8249 & 0.8505 \\
$S_{B,L,S}^{rG,rG,rG}$ & \underline{0.9428} & \underline{0.9215} & 0.8866 & \multicolumn{1}{c|}{\underline{0.9170}} & \underline{0.8583} & 0.8790 & \underline{0.8268} & \underline{0.8547} \\
$S_{B,L,S}^{R,R,R}$ & 0.9405 & 0.9186 & 0.8783 & \multicolumn{1}{c|}{0.9124} & 0.8531 & \underline{0.8809} & 0.8196 & 0.8512 \\
All & \textbf{0.9432} & \textbf{0.9229} & 0.8861 & \multicolumn{1}{c|}{\textbf{0.9174}} & \textbf{0.8663} & \textbf{0.8839} & \textbf{0.8291} & \textbf{0.8598} \\ \hline
\end{tabular}
\end{table}

\begin{table}[]
\centering
\caption{Results of the validation set of our best solution, i.e., "All" $S_{B,L,S, B, L, S,B,L,S}^{G,G,G,rG,rG,rG,R,R,R}$ with threshold of 250, 150, 100, and the winners of 2021 and 2022. The best results are in \textbf{bold} and the second best \underline{underlined}}
\label{tab:our_winner2022}
\begin{tabular}{|c|cccccccc|}
\hline
\textbf{} & \multicolumn{4}{c}{\textbf{Legacy DSC}} & \multicolumn{4}{c|}{\textbf{Legacy HD95}} \\ \hline
Solutions & ET & TC & WT & \multicolumn{1}{c|}{Mean} & ET & TC & WT & Mean \\ \hline
All & 0.8464 & 0.8769 & \textbf{0.9294} & \multicolumn{1}{c|}{0.8842} & 17.81 & 11.12 & 4.26 & 11.06 \\ 

$S_{B,L,S}^{rG,rG,rG}$ & \underline{0.8484} & \textbf{0.8781} & \underline{0.9286} & \multicolumn{1}{c|}{\underline{0.8850}} & \underline{17.75} & 11.08 & 4.19 & 11.00 \\

$S_{B,L,S}^{G,G,G}$ & \textbf{0.8515} & 0.8761 & 0.9283 & \multicolumn{1}{c|}{\textbf{0.8853}} & 19.32 & 12.70 & 4.27 & 12.10 \\

$S_{B,L,S}^{R,R,R}$ & 0.8318 & 0.8719 & 0.9279 & \multicolumn{1}{c|}{0.8772} & 21.36 & 13.04 & 4.61 & 13.00 \\ 

2022 \cite{zeineldin2022multimodal} & 0.8438 & 0.8753 & 0.9271 & \multicolumn{1}{c|}{0.8821} & \textbf{17.50} & \textbf{7.53} & \underline{3.60} & \textbf{9.54} \\

2021 \cite{luu2021extending} & 0.8451 & \textbf{0.8781} & 0.9275 & \multicolumn{1}{c|}{0.8836} & 20.73 & \underline{7.62} & \textbf{3.47} & \underline{10.61} \\ \hline
\end{tabular}
\end{table}

\begin{table}[]
\centering
\caption{Validation set results computed by the validation platform (new metric). The values between parentheses are the threshold value used for (WT, TC, ET) respectively. The best results are in \textbf{bold} and the second best \underline{underlined}. The "All" is defined as $S_{B,L,S,B,L,S,B,L,S}^{G,G,G,rG,rG,rG,R,R,R}$}
\label{tab:validation}
\begin{tabular}{|c|ccccccccccc|}
\hline

\textbf{} & \multicolumn{3}{c}{\textbf{Thresholds}} & \multicolumn{4}{c}{\textbf{DSC}} & \multicolumn{4}{c|}{\textbf{HD95}} \\ \hline
Solutions & WT & TC & \multicolumn{1}{c|}{ET} & WT & TC & ET & \multicolumn{1}{c|}{Mean} & WT & TC & ET & Mean \\ \hline
All & 250 & 150 & \multicolumn{1}{c|}{100} & \underline{0.9005} & \underline{0.8673} & 0.8509 & \multicolumn{1}{c|}{\underline{0.8729}} & \underline{14.940} & \underline{14.467} & 17.699 & \underline{15.702} \\

All & 1450 & 150 & \multicolumn{1}{c|}{100} & \textbf{0.9101} & \underline{0.8673} & 0.8509 & \multicolumn{1}{c|}{\textbf{0.8761}} & \textbf{11.113} & \underline{14.467} & 17.699 & \textbf{14.426} \\
$S_{B,L,S}^{G,G,G}$ & 100 & 50 & \multicolumn{1}{c|}{100} & 0.8867 & 0.8575 & \textbf{0.8537} & \multicolumn{1}{c|}{0.8660} & 19.322 & 18.807 & \textbf{17.321} & 18.483 \\
$S_{B,L,S}^{G,G,G}$ &300 & 200 & \multicolumn{1}{c|}{200} & 0.8969 & 0.8660 & \underline{0.8528} & \multicolumn{1}{c|}{0.8719} & 15.798 & 16.397 & 19.470 & 17.222 \\
$S_{B,L,S}^{rG,rG,rG}$ & 100 & 50 & \multicolumn{1}{c|}{50} & 0.8918 & 0.8565 & 0.8347 & \multicolumn{1}{c|}{0.8610} & 17.201 & 19.352 & 23.880 & 20.144 \\
$S_{B,L,S}^{rG,rG,rG}$ & 250 & 150 & \multicolumn{1}{c|}{100} &  0.8972 & \textbf{0.8686} & 0.8527 & \multicolumn{1}{c|}{0.8728} &  15.510 & \textbf{14.420} & \underline{17.643} & 15.858 \\
$S_{B,L,S}^{R,R,R}$ & 100 & 50 & \multicolumn{1}{c|}{50} & 0.8874 & 0.8564 & 0.8269 & \multicolumn{1}{c|}{0.8569} & 19.637 & 19.090 & 23.794 & 20.840 \\
$S_{B,L,S}^{R,R,R}$ & 250 & 150 & \multicolumn{1}{c|}{100} & 0.8990 & 0.8606 & 0.8350 & \multicolumn{1}{c|}{0.8648} & 15.121 & 18.275 & 21.049 & 18.148 \\
\hline
\end{tabular}
\end{table}

\subsection{Qualitative results}

\begin{figure}[h!]
     \centering
     \begin{subfigure}[b]{0.19\textwidth}
         \centering
         \includegraphics[width=\textwidth]{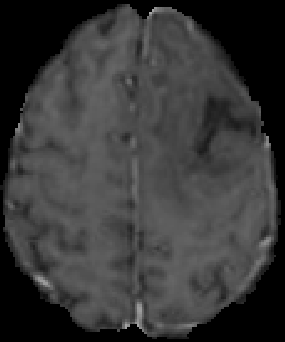}
         \caption{T1CE}
         \label{fig:1774_cT1}
     \end{subfigure}
     \hfill
     \begin{subfigure}[b]{0.19\textwidth}
         \centering
         \includegraphics[width=\textwidth]{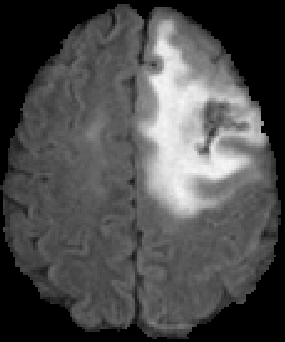}
         \caption{FLAIR}
         \label{fig:1774_flair}
     \end{subfigure}
     \hfill
     \begin{subfigure}[b]{0.19\textwidth}
         \centering
         \includegraphics[width=\textwidth]{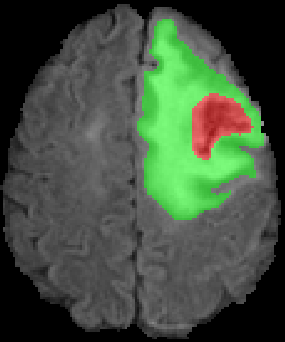}
         \caption{Prediction}
         \label{fig:1774_ct1_label}
     \end{subfigure}
     
     \begin{subfigure}[b]{0.19\textwidth}
         \centering
         \includegraphics[width=\textwidth]{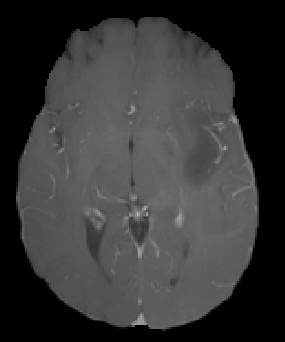}
         \caption{T1CE}
         \label{fig:521-001_T1CE}
     \end{subfigure}
     \hfill
     \begin{subfigure}[b]{0.19\textwidth}
         \centering
         \includegraphics[width=\textwidth]{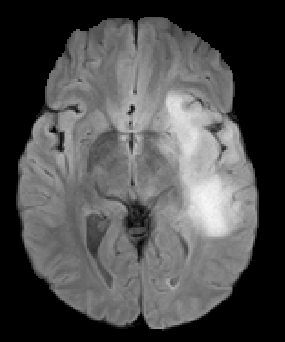}
         \caption{FLAIR}
         \label{fig:521-001_FLAIR}
     \end{subfigure}
     \hfill
     \begin{subfigure}[b]{0.19\textwidth}
         \centering
         \includegraphics[width=\textwidth]{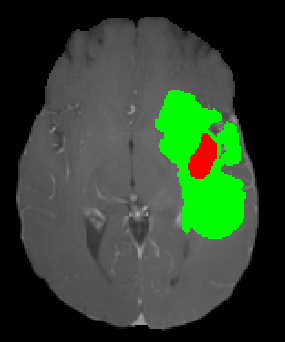}
         \caption{Prediction}
         \label{fig:521-001_T1CE_label}
     \end{subfigure}
    
     \begin{subfigure}[b]{0.19\textwidth}
         \centering
         \includegraphics[width=\textwidth]{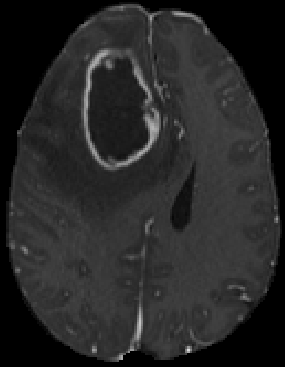}
         \caption{T1CE}
         \label{fig:190_ct1_label}
     \end{subfigure}
     \hfill
     \begin{subfigure}[b]{0.19\textwidth}
         \centering
         \includegraphics[width=\textwidth]{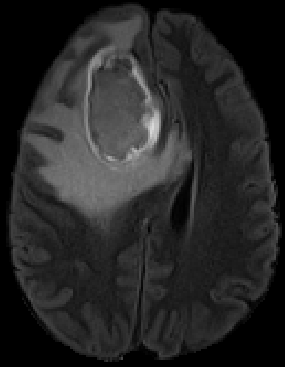}
         \caption{FLAIR}
         \label{fig:190_ct1_label}
     \end{subfigure}
     \hfill
     \begin{subfigure}[b]{0.19\textwidth}
         \centering
         \includegraphics[width=\textwidth]{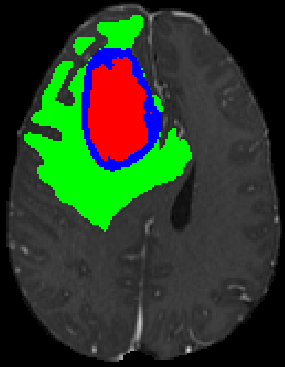}
         \caption{Prediction}
         \label{fig:190_ct1_label}
     \end{subfigure}
    
    \caption{Samples of the validation set segmented by our best ensemble model. Columns display the T1CE, T2-FLAIR, and the overlay of our
predicted segmentation on the T1CE scan. Red corresponds to NCR, green ED, and blue ET. 
The first row presents the case with ID 01774-001, the second row the worse prediction (ID 00521-001), and the third row the best prediction (ID 00190-00).}
    \label{fig:validation_predictions}
\end{figure}

Figure \ref{fig:validation_predictions} presents the predicted segmentations of cases 01774-000, 00521-001 and 00190-000. In the first two cases our models performed poorly (Figure \ref{fig:validation_predictions} first and second rows). The ET is not detected by our solutions, which leads to a value 0 for this label. This could be due to the quality of the acquired scans, as this strongly influences the performance of deep learning models. For future improvements, more synthetic data can be generated with the specific cases of poorer scans.
The last row of Figure \ref{fig:validation_predictions} shows a case with an almost perfect segmentation.
Our solution archived DSC above 0.9 for most of the validation cases.

\section{Discussion}

From Table \ref{tab:5-fold} we can conclude that the best solution is the ensemble (average of the probability maps of each model and rounded to an integer). However, if we compare the old results with the results of the new metric, we can also deduce that our solution is heavily penalised by the existence of FP and FN. Therefore, post-processing based on a threshold is performed to remove some small tumours that are detected but are not actually tumours. For this purpose, several values were tested for each region (WT, TC and ET). It was found that the best thresholds for the training set are $WT_{250}TC_{100}ET_{50}$ (for DSC) and $WT_{250}TC_{50}ET_{50}$ (for HD96). However, for the test in the validation set, the best values are $WT_{250}TC_{150}ET_{100}$ (for both DSC and HD95). 

Table \ref{tab:our_winner2022} compares the legacy results of our solution  with the winners of 2021 and 2022. The results are very similar. However, our solution archives better dice scores but worse HD95, specially for the tumour core. This can be related with the value of threshold used or with the fact that, since our models were trained with a larger variety of data, the predictions might have larger variety, which increases the HD95 distance. Regarding only our solutions, it can be seen that the ensemble $S_{B,L,S}^{G,G,G}$ archives the best mean DSC, and that $S_{B,L,S}^{rG,rG,rG}$ the best HD95. Perhaps, the ensemble $S_{B,L,S,B,L,S}^{G,G,G,rG,rG,rG}$ would produce better results than our final submission, however this was not tested due to lack of time.

Several other combinations of ensembles and thresholds were tested in the validation set, and it was consistently found that the validation set required larger thresholds than the training set. It was also found that disabling the data augmentation option of the nnUNet during inference yielded better results in the validation set, which also allows the use of more models for the ensemble, as "fast" inference is up to 8 times faster than normal inference.

Table \ref{tab:validation} shows the extension of the influence of large values of threshold. Two different thresholds are given for each ensemble. The first is the threshold with the best results in the training set (using the ranking system) and the second adjustments after analysing the results provided by the platform after running the first. It can be seen that larger thresholds are required for the validation set compared to the training set. However, it is important to note that we cannot increase the threshold too much, as this could increase the number of FNs. The second row of the table shows the best overall results, however a value of 1450 voxels for the threshold of the WT might be too high for the test group. Therefore, the first solution is chosen for the test phase.

We can also conclude that the data augmentation with GliGAN gives the better DSC and HD95 for the ET. Furthermore, the ensemble of all available models gives the best solution using the ranking system (as explained in \cite{isensee2021nnu_brats}) both in the training and validation set.

The other ensemble strategies were also tested but produced worse results than regular averaging, so they are not presented here. While these strategies have the potential to improve the results, since the focus of our solution is on using synthetic data, most of the effort has been spent on improving the synthetic data, rather than improving the ensemble strategy.

The major limitation of the GliGAN is that it is only capable of generate tumours with dimensions 96×96×96. This would be solved by using a generator with greater input size. Since the generator is a Swin UNETR, an input image greater than 96×96×96 could be used \cite{tang2022self}, but this was not tested.

For future improvements of our solution: The use of the ensemble $S_{B,L,S,B,L,S}^{G,G,G,rG,rG,rG}$ should be tested with the validation set; Identify which cases yields worse results and produce more synthetic data based on those cases; Use the synthetic data to train other networks.

\subsubsection{Acknowledgements} André Ferreira thanks the Fundação para a Ciência e Tecnologia (FCT) Portugal for the grant 2022.11928.BD. This work has been supported by FCT within the R\&D Units Project Scope: UIDB/00319/2020 and this work received funding from enFaced (FWF KLI 678), enFaced 2.0 (FWF KLI 1044) and KITE (Plattform für KI-Translation Essen, EFRE-0801977) from the REACT-EU initiative (\url{https://kite.ikim.nrw/}).

%
%
%

\bibliographystyle{splncs04}
\bibliography{bib}

\end{document}